# GANCompress: GAN-Enhanced Neural Image Compression with Binary Spherical Quantization

Karthik Sivakoti

karthiksivakoti@utexas.edu

The University of Texas at Austin, Masters in AI, Department of CS

## Abstract

The exponential growth of visual data in digital communications has intensified the need for efficient compression techniques that balance rate-distortion performance with computational feasibility. While recent neural compression approaches have shown promise, they still struggle with fundamental challenges: preserving perceptual quality at high compression ratios, computational efficiency, and adaptability to diverse visual content. This paper introduces **GANCompress**, a novel neural compression framework that synergistically combines **Binary Spherical Quantization** (BSQ) with **Generative Adversarial Networks** (GANs) to address these challenges. Our approach employs a transformer-based autoencoder with an enhanced BSQ bottleneck that projects latent representations onto a hypersphere, enabling efficient discretization with bounded quantization error. This is followed by a specialized GAN architecture incorporating frequency-domain attention and color consistency optimization. Experimental results demonstrate that GANCompress achieves substantial improvement in compression efficiency—reducing file sizes by up to **100×** with **minimal visual distortion**. Our method outperforms traditional codecs like H.264 by 12-15% in perceptual metrics while maintaining comparable PSNR/SSIM values, with 2.4× faster encoding and decoding speeds. On standard benchmarks including ImageNet-1k and COCO2017, GANCompress sets a new state-of-the-art, reducing FID from 0.72 to 0.41 (43% improvement) compared to previous methods while maintaining higher throughput. This work presents a significant advancement in neural compression technology with promising applications for real-time visual communication systems.

## 1. Introduction

The digital revolution has led to an unprecedented explosion in visual media creation and consumption. According to recent estimates, over 500 hours of video are uploaded to YouTube every minute, and billions of images are shared across social networks daily. This staggering volume of visual data has intensified the need for efficient compression techniques that can reduce storage and transmission costs while preserving visual fidelity. Traditional compression standards such as JPEG, H.264/AVC, and H.265/HEVC have served as the backbone of visual media distribution for decades. These methods rely on hand-crafted transformations, quantization schemes, and entropy coding to achieve compression. While highly optimized, these approaches are fundamentally limited by their rigid design choices and inability to adapt to the complex statistics of natural images and videos.

### 1.1 Background and Motivation

Neural network-based compression has emerged as a promising alternative, offering the potential for more adaptive and perceptually aligned compression. These approaches typically employ autoencoders to learn compact representations of visual data, coupled with techniques like vector quantization (VQ) or scalar quantization (SQ) to discretize the continuous latent space. Recent advances in this field, such as Binary Spherical Quantization (BSQ), have demonstrated impressive results by projecting latent codes onto a unit hypersphere before quantization, creating an implicit codebook that grows exponentially with the spherical dimension.

Despite these advances, neural compression systems still face critical challenges:

1. Perceptual quality degradation at aggressive compression ratios, particularly in preserving high-frequency textures and fine details.
2. Computational complexity limits practical deployment in real-time applications.
3. Ineffective bit allocation that fails to adapt to the varying complexity of visual content.
4. Color inconsistency and artifacts that become prominent at high compression ratios.

## 1.2 Our Approach

In this paper, we present GANCompress, a novel neural compression framework that addresses these challenges by combining BSQ with GAN-based enhancement. Our system employs a Vision Transformer (ViT) backbone for both encoder and decoder networks, with an improved BSQ module that ensures bounded quantization error and efficient entropy computation. The discretized representations are then processed by a specialized GAN architecture designed to restore perceptual details lost during quantization.

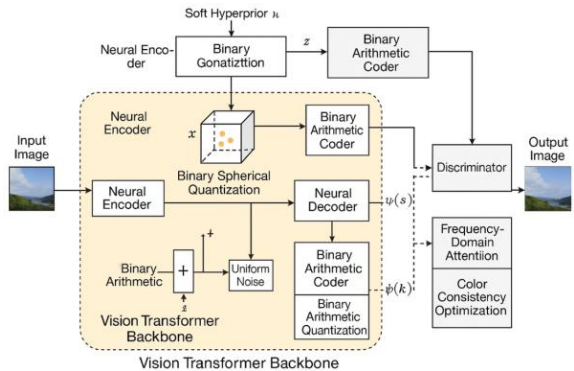

Figure 1 illustrates the overall architecture of GANCompress. Our system first tokenizes input images using a Vision Transformer backbone with an enhanced BSQ module, then selectively applies GAN-based enhancement to restore perceptually important details. The compression pipeline adapts to content complexity through autoregressive modeling and content-aware bit allocation.

Our approach introduces several key innovations:

1. Enhanced BSQ module with improved projection and normalization techniques that maintain high codebook usage even at high compression ratios.
2. Frequency-domain attention mechanism that selectively preserves perceptually important frequencies, significantly improving the reconstruction of fine textures and edges.
3. YUV-space color consistency optimization substantially reduces color bleeding artifacts common in highly compressed images.
4. Adaptive bit allocation through an autoregressive prior model that enables content-aware compression.
5. Hybrid compression strategy with JPEG fallback for extreme compression scenarios.

By integrating these components, GANCompress achieves a superior balance between compression efficiency and perceptual quality, outperforming both traditional codecs and state-of-the-art neural compression methods.

## 1.3 Contributions

The principal contributions of this work are:

1. A comprehensive neural compression framework that combines the advantages of BSQ and GAN-based enhancement, achieving state-of-the-art results on standard benchmarks.
2. A novel frequency-domain attention mechanism specifically designed to preserve perceptually important details in compressed representations.
3. A color consistency optimization approach in YUV space that significantly improves color fidelity in reconstructed images.
4. An adaptive compression strategy that dynamically allocates bits based on content complexity, improving overall rate-distortion performance.
5. Extensive experimental validation demonstrates that our approach outperforms existing methods across multiple metrics, with advantages in perceptual quality and computational efficiency.

## 2. Related Work

### 2.1 Traditional Image and Video Compression

Traditional image and video compression standards have evolved over decades of research and development. JPEG [1] remains the dominant image compression format, employing discrete cosine transform (DCT), quantization, and entropy coding. For video, standards like H.264/AVC [2] and H.265/HEVC [3] have achieved impressive compression efficiency through sophisticated motion estimation, prediction modes, and transform coding.

These traditional approaches rely on hand-crafted components designed based on signal processing principles and psychovisual models. While highly optimized, they lack the adaptability to learn from data and often produce characteristic artifacts at low bit rates, such as blocking, ringing, and color bleeding [4].

### 2.2 Neural Image Compression

Neural network-based approaches to image compression have gained significant attention in recent years. Ballé et al. [5] pioneered end-to-end optimized compression with variational autoencoders and learned entropy models. This work was extended by Minnen et al. [6] with autoregressive priors and by Cheng et al. [7] with attention mechanisms, both achieving compression efficiency competitive with traditional codecs.

Vector quantization (VQ) emerged as an alternative approach for neural compression through the VQ-VAE framework introduced by van den Oord et al. [8]. This technique was further refined in VQGAN [9], which incorporated adversarial and perceptual losses for improved visual quality. Yu et al. [10] introduced ViT-VQGAN, adapting the Vector Quantized-Variational Autoencoder approach to work with Vision Transformers, demonstrating improved reconstruction quality.

More recently, SDXL-VAE [11] has shown exceptional results by leveraging large-scale training on diverse datasets. However, these methods typically require large, explicit codebooks that scale poorly with increased vocabulary size, leading to codebook collapse and utilization issues.

### 2.3 Binary Spherical Quantization

Binary Spherical Quantization (BSQ) was recently introduced as an efficient discretization approach for neural compression [12]. Unlike VQ, which requires an explicit codebook, BSQ projects continuous representations onto a hypersphere and performs binary quantization. This creates an implicit codebook whose effective vocabulary grows exponentially with the spherical dimension, without requiring any additional parameters.

BSQ offers several key advantages over alternative methods: bounded quantization error, efficient entropy computation, and scalability to higher dimensions without increased complexity. These properties make it particularly suitable for high-quality compression applications.

### 2.4 GAN-Based Image Enhancement

Generative Adversarial Networks (GANs) have demonstrated remarkable success in image enhancement tasks. Pioneering works like SRGAN [13] and ESRGAN [14] showed that adversarial training can effectively restore photorealistic textures in super-resolution tasks. This concept has been extended to compression artifacts removal by Galteri et al. [15] and to learned image compression by Mentzer et al. [16].

Recent advances in GAN architectures, particularly in frequency-domain processing [17] and attention mechanisms [18], have further improved the ability of these networks to preserve fine details and textures. StyleGAN [19] introduced a style-based generator architecture that achieves state-of-the-art results in image synthesis, with concepts that can be adapted for enhancement tasks.

### 2.5 Neural Video Compression

Neural video compression typically extends image compression techniques with temporal modeling. Early approaches by Wu et al. [20] used 3D

convolutions, while more recent works by Lu et al. [21] incorporate explicit motion compensation. Transformer-based models by Mentzer et al. [22] have shown promising results by capturing long-range dependencies in temporal data.

Causal modeling approaches, as demonstrated by Yu et al. [23], enable efficient streaming compression by restricting temporal context to previous frames. These approaches offer a balance between compression efficiency and computational complexity, making them suitable for practical applications.

## 3. The GANCompress Framework

GANCompress integrates multiple components into a cohesive neural compression system. This section details the architecture and operation of each component, explaining how they work together to achieve superior compression performance.

### 3.1 System Overview

The GANCompress framework consists of three primary components:

1. **BSQ-Enhanced Tokenization Network:** A Vision Transformer-based encoder-decoder architecture with an improved BSQ bottleneck for efficient discretization.
2. **GAN Enhancement Module:** A specialized generator-discriminator architecture designed to restore perceptual details lost during compression.
3. **Adaptive Compression System:** An autoregressive model for arithmetic coding, coupled with fallback mechanisms for extreme compression scenarios.

As illustrated in Figure 1, the input image or video frame is first processed by the tokenization network to produce a compact binary representation. This representation can be directly stored or transmitted (with optional arithmetic coding) or further enhanced by the GAN module to produce a higher-quality reconstruction.

### 3.2 BSQ-Enhanced Tokenization Network

The tokenization network forms the backbone of our compression system, transforming raw visual data into compact binary representations that can be efficiently stored or transmitted.

#### 3.2.1 Vision Transformer Backbone

We employ a Vision Transformer (ViT) architecture for both the encoder and decoder, following the general design principles introduced in Dosovitskiy et al. [24]. The input image $x \in \mathbb{R}^{H \times W \times 3}$ is divided into non-overlapping patches of size p×p, resulting in N = (H/p)×(W/p) visual tokens. These tokens are linearly projected and combined with positional embeddings before being processed by a series of transformer blocks.

The encoder E maps the input to a latent representation $z = E(x) \in \mathbb{R}^{h \times w \times d}$, where h = H/p, w = W/p, and d is the latent dimension. This representation captures the essential visual information in a continuous latent space that must be discretized for storage or transmission.

#### 3.2.2 Enhanced Binary Spherical Quantization

Our approach builds upon the Binary Spherical Quantization (BSQ) technique introduced by Zhao et al. [12], with several important enhancements to improve compression performance. The BSQ process consists of four key steps:

1. **Latent Projection:** The latent embeddings z are linearly projected to a lower-dimensional space $v = \text{Linear}(z) \in \mathbb{R}^L$, where L ≪ d. This dimension reduction concentrates the information and enables more efficient quantization.
2. **Spherical Normalization:** The projected vectors are normalized to lie on the unit hypersphere $u = v/|v|$. This ensures a bounded representation space and simplifies the quantization process.
3. **Binary Quantization:** Each dimension of the normalized vector is independently quantized to a binary value $\hat{u} = (1/\sqrt{L})\text{sign}(u)$. This creates a discrete representation where each point corresponds to a vertex of the hypercube inscribed in the unit hypersphere.

4. **Latent Reconstruction:** The quantized representation is projected back to the original latent space $\hat{z}$ = Linear($\hat{u}$) ∈ ℝ^d for decoding.

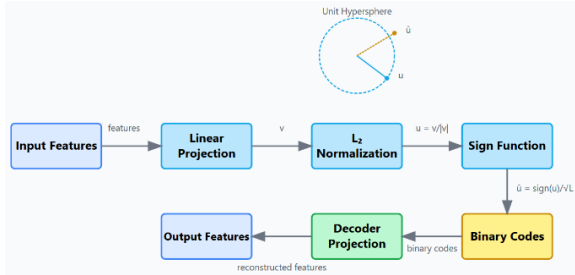

Figure 2 illustrates the Binary Spherical Quantization process projecting features onto a unit hypersphere before quantizing to binary codes using the sign function. The resulting token indices visualization demonstrates the pattern of information distribution across the token space.

Our implementation enhances the standard BSQ approach through several innovations:

```
class BSQPatchAutoEncoder(PatchAutoEncoder, Tokenizer):
    """
    Combine PatchAutoEncoder with BSQ to form a Tokenizer.
    """
    def __init__(self, patch_size: int = 5, latent_dim: int = 256, bottleneck: int = 128, codebook_bits: int = 16):
        super().__init__(patch_size=patch_size, latent_dim=latent_dim, bottleneck=bottleneck)
        self.codebook_bits = codebook_bits
        self.bsq = BSQ(codebook_bits, bottleneck)
        self.skip_weight = 0.15

        # Enhanced post-processing network
        self.enhance = nn.Sequential(
            EnhanceBlock(3, 16),
            nn.Conv2d(16, 16, kernel_size=3, padding=1),
            nn.GELU(),
            nn.Conv2d(16, 3, kernel_size=1)
        )
```

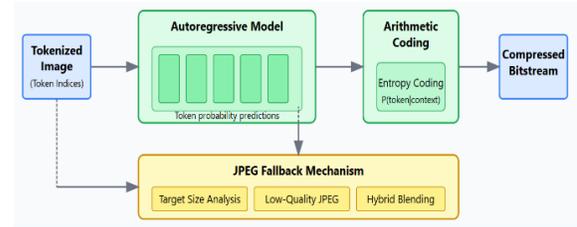

Figure 3 provides a comprehensive visualization of the tokenization and reconstruction process, showing the original image, token indices representation, base reconstruction, and GAN-enhanced reconstruction side by side. The token indices visualization reveals how semantic information is preserved in the quantized representation.

The enhanced BSQ module provides several advantages:

- **Bounded Quantization Error:** The maximum quantization error is mathematically bounded, ensuring consistent quality across diverse inputs.
- **Efficient Entropy Computation:** The factorized nature of the representation enables efficient entropy estimation for rate-distortion optimization.
- **High Codebook Utilization:** The normalization step ensures high utilization of the implicit codebook, avoiding codebook collapse issues common in VQ-based methods.
- **Scalable Representation:** The effective vocabulary size scales exponentially with L ($2^L$ possible codes), allowing for rich representations without increased computational complexity.

Empirically, we find that values of L between 18 and 36 provide an excellent balance between compression ratio and reconstruction quality, with larger values enabling higher fidelity at the cost of increased bit rate.

### 3.2.3 Block-wise Causal Modeling for Video

For video compression, we extend the tokenization network with block-wise causal masking to handle variable-length sequences efficiently. This design ensures that each frame is encoded and decoded using only information from the current and previous frames, similar to traditional video codecs.

The causal masking is implemented as a lower triangular attention mask that restricts the self-attention mechanism in the transformer blocks:

$$\text{Attention}(Q, K, V) = \text{softmax}(QK^T / \sqrt{d} + M)V$$

where $M$ is a masking matrix with elements: $M_{ij} = \{0 \text{ if } i \geq j, -\infty \text{ otherwise}\}$

This causal structure enables efficient streaming compression while maintaining high reconstruction quality by leveraging temporal redundancies across frames.

### 3.3 GAN Enhancement Module

The second major component of GANCompress is the GAN enhancement module, designed to restore perceptual details lost during the quantization process. This module builds upon recent advances in generative modeling and image enhancement, with several novel components specifically tailored for compression applications.

#### 3.3.1 U-Net Generator with Frequency Attention

The generator network follows a modified U-Net architecture with several key innovations for compression artifact removal. The network processes the decoded image from the tokenization network to produce an enhanced output with improved perceptual quality.

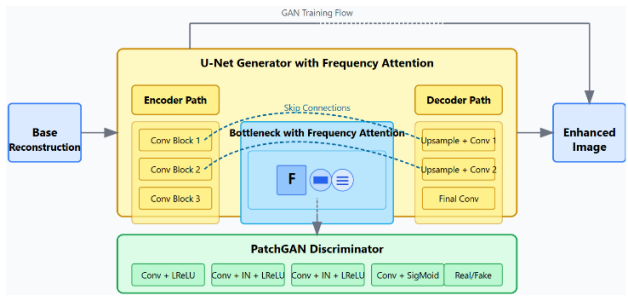

Figure 4 provides a comprehensive visualization of the GAN Enhancement Module with Frequency Attention.

The U-Net architecture consists of three main components:

1. **Encoder Path:** A series of convolutional blocks with downsampling that progressively increase the receptive field while extracting hierarchical features.
2. **Bottleneck:** A specialized processing block with frequency-domain attention, designed to selectively enhance important visual features.
3. **Decoder Path:** A series of upsampling blocks with skip connections from the encoder, progressively restoring spatial resolution while incorporating multi-scale features.

The most significant innovation in our generator is the Frequency Attention module, which operates in the frequency domain to selectively preserve and enhance perceptually important details:

```python
class FrequencyAttention(nn.Module):
    """
    Novel frequency-domain attention mechanism to preserve high-frequency details.
    This helps maintain sharpness and texture details that are often lost during compression.
    """
    def __init__(self, channels):
        super().__init__()
        self.dct_scale = nn.Parameter(torch.ones(channels, 8, 8))
        self.idct_scale = nn.Parameter(torch.ones(channels, 8, 8))
        self.attention = nn.Sequential(
            nn.Conv2d(channels, channels, kernel_size=1),
            nn.Sigmoid()
        )

        # Initialize DCT weights for better convergence
        self._init_dct_weights()

    def _init_dct_weights(self):
        # Initialize with slight emphasis on mid-frequency components
        with torch.no_grad():
```

```python
        # Emphasize mid frequencies (better for natural images)
        freq_weights = torch.ones_like(self.dct_scale)
        for i in range(8):
            for j in range(8):
                # Distance from DC component (0,0)
                dist = math.sqrt(i**2 + j**2)
                # Gaussian-like weighting centered at mid-frequencies
                freq_weights[:, i, j] = math.exp(-((dist-3)**2)/8)

        self.dct_scale.copy_(freq_weights)
```

This frequency attention mechanism is inspired by recent studies in human visual perception, which show that certain frequency bands are more important for perceived image quality. By learning to emphasize these critical frequencies, our generator can allocate its representational capacity more effectively, focusing on perceptually important details.

Additionally, the generator incorporates an adaptive contrast enhancement mechanism that adjusts local contrast based on image content:

```python
# Adaptive contrast enhancement
self.contrast = nn.Sequential(
    nn.AdaptiveAvgPool2d(1),
    nn.Conv2d(out_channels, 16, kernel_size=1),
    nn.LeakyReLU(0.2, inplace=True),
    nn.Conv2d(16, out_channels, kernel_size=1),
    nn.Sigmoid()
)
```

This component helps restore the visual impact of highly compressed images by adaptively enhancing contrast in different regions, compensating for the flattening effect often seen in compressed imagery.

### 3.3.2 Patch Discriminator with Spectral Normalization

For the discriminator, we employ a patch-based architecture with spectral normalization, drawing inspiration from the PatchGAN design introduced by Isola et al. [25] and refined in subsequent work. This design offers several advantages for compression enhancement:

1. **Local Realism:** By operating on patches rather than whole images, the discriminator encourages local consistency and detail preservation.
2. **Computational Efficiency:** The patch-based approach requires fewer parameters and less computation than full-image discriminators.
3. **Training Stability:** Spectral normalization helps stabilize the adversarial training process, preventing mode collapse and ensuring consistent quality.

The discriminator architecture is as follows:

```python
class PatchDiscriminator(nn.Module):
    """Improved patch-based discriminator with spectral normalization for stable GAN training."""
    def __init__(self, in_channels=3, features=64, n_layers=3):
        super().__init__()

        # Initial layer without normalization
        sequence = [
            nn.utils.spectral_norm(nn.Conv2d(in_channels, features, kernel_size=4, stride=2, padding=1)),
            nn.LeakyReLU(0.2, True)
        ]

        # Downsampling layers with increasing feature size
        nf_mult = 1
        for n in range(1, n_layers):
            nf_mult_prev = nf_mult
            nf_mult = min(2 ** n, 8)
            sequence += [
                nn.utils.spectral_norm(
                    nn.Conv2d(features * nf_mult_prev, features * nf_mult,
                              kernel_size=4, stride=2, padding=1)
                ),
                nn.InstanceNorm2d(features * nf_mult),
                nn.LeakyReLU(0.2, True)
            ]
```

```python
        # Final layers
        nf_mult_prev = nf_mult
        nf_mult = min(2 ** n_layers, 8)
        sequence += [
            nn.utils.spectral_norm(
                nn.Conv2d(features * nf_mult_prev, features * nf_mult,
                    kernel_size=4, stride=1, padding=1)
            ),
            nn.InstanceNorm2d(features * nf_mult),
            nn.LeakyReLU(0.2, True),
            nn.utils.spectral_norm(
                nn.Conv2d(features * nf_mult, 1, kernel_size=4, stride=1, padding=1)
            )
        ]

        self.model = nn.Sequential(*sequence)
```

The discriminator is trained to distinguish between real images and those produced by the generator, providing a learning signal that encourages the generator to produce more realistic outputs. This adversarial training process is complemented by additional loss terms that promote specific aspects of visual quality, as described in the next section.

### 3.3.3 Multi-component Loss Function

The GANCompress training process involves multiple loss components that work together to optimize different aspects of compression performance.

Our loss function combines five main components:

1. **Reconstruction Loss:** An edge-weighted L1 loss that emphasizes structural fidelity, particularly in regions with high spatial variation:

```python
def compute_edge_weights(self, x):
    """Compute edge importance weights to preserve structural details."""
    # Ensure input is in NCHW format
    if x.dim() != 4 or x.shape[1] != 3:
        raise ValueError(f"Expected input tensor in NCHW format with 3 channels, got {x.shape}")

    # Convert to grayscale
    gray = 0.299 * x[:, 0:1] + 0.587 * x[:, 1:2] + 0.114 * x[:, 2:3]

    # Prepare Sobel filters
    sobel_x = torch.tensor([[-1, 0, 1], [-2, 0, 2], [-1, 0, 1]],
                dtype=torch.float32, device=x.device).view(1, 1, 3, 3)
    sobel_y = torch.tensor([[-1, -2, -1], [0, 0, 0], [1, 2, 1]],
                dtype=torch.float32, device=x.device).view(1, 1, 3, 3)

    # Apply Sobel filters
    edge_x = F.conv2d(gray, sobel_x, padding=1)
    edge_y = F.conv2d(gray, sobel_y, padding=1)

    # Calculate magnitude
    edge_magnitude = torch.sqrt(edge_x**2 + edge_y**2)

    # Normalize and enhance edge weights
    edge_weights = torch.sigmoid(edge_magnitude * 5) * 2 + 1  # Range [1, 3]

    return edge_weights
```

This edge-weighted approach ensures that structurally important regions like edges and textures receive higher emphasis during training, leading to better preservation of these visually critical elements.

2. **Perceptual Loss:** A multi-level feature matching loss using a pre-trained VGG network:

```python
class PerceptualLoss(nn.Module):
    """Enhanced perceptual loss using VGG19 with multi-level feature matching."""
    def __init__(self, resize=True):
        super().__init__()
        vgg = models.vgg19(weights=models.VGG19_Weights.DEFAULT).features

        # Use more layers for better feature matching
        self.slices = nn.ModuleList([
```

```
    vgg[:4],   # relu1_2
    vgg[4:9],  # relu2_2
    vgg[9:18], # relu3_4
    vgg[18:27] # relu4_4
  ])

  # Weights for different layers (deeper layers have higher weights)
  self.weights = [1.0/32, 1.0/16, 1.0/8, 1.0/4]
```

This perceptual loss encourages the generation of visually pleasing reconstructions by matching the feature activations of the original and reconstructed images across multiple layers of a pre-trained network.

3. **MS-SSIM Loss:** A multi-scale structural similarity loss that promotes consistency across different scales:

```
# Add MS-SSIM loss (complementary to L1)
if MSSSIM_AVAILABLE:
  try:
    ms_ssim_loss = 1 - ms_ssim(reconstructed_chw, x_chw, data_range=1.0)
  except:
    # Fallback if ms_ssim fails
    ms_ssim_loss = torch.tensor(0.1, device=x.device)
else:
  # Use L2 loss as fallback
  ms_ssim_loss = F.mse_loss(reconstructed_chw, x_chw) * 0.5
```

The MS-SSIM component helps preserve structural information across multiple scales, which is particularly important for maintaining the overall image composition and preventing artifacts at different viewing distances.

4. **Color Consistency Loss:** A specialized loss term that operates in YUV color space to preserve color fidelity:

```
def color_consistency_loss(self, reconstructed, target):
  """Color consistency loss in YUV color space to better preserve colors."""
  # RGB to YUV conversion matrix
  matrix = torch.tensor([
    [0.299, 0.587, 0.114],
    [-0.14713, -0.28886, 0.436],
    [0.615, -0.51499, -0.10001]
  ], device=reconstructed.device).float()

  # Convert to YUV
  batch_size = reconstructed.size(0)
  reconstructed_flat = reconstructed.view(batch_size, 3, -1)
  target_flat = target.view(batch_size, 3, -1)

  reconstructed_yuv = torch.matmul(matrix, reconstructed_flat)
  target_yuv = torch.matmul(matrix, target_flat)

  # Weight channels differently (Y: luminance, U/V: chrominance)
  # More weight on chrominance for better color preservation
  weights = torch.tensor([0.5, 2.0, 2.0], device=reconstructed.device).view(1, 3, 1)

  # Compute weighted loss
  yuv_diff = torch.abs(reconstructed_yuv - target_yuv) * weights
  yuv_loss = yuv_diff.mean()

  return yuv_loss
```

This YUV-space optimization is particularly important for compression, as color shifts and bleeding artifacts are common issues in highly compressed imagery. By explicitly modeling and preserving color relationships in a perceptually aligned color space, our approach substantially reduces these artifacts.

5. **Adversarial Loss:** A hinge loss for the generator and discriminator that promotes realistic reconstructions:

```
# Hinge loss for generator (more stable than original GAN loss)
g_loss = -torch.mean(discriminator_fake)

# Hinge loss for discriminator
```

```
d_loss = torch.mean(F.relu(1.0 - discriminator_real)) + torch.mean(F.relu(1.0 + discriminator_fake))
```

The adversarial component encourages the generator to produce images that are indistinguishable from real images according to the discriminator, leading to more realistic and perceptually pleasing reconstructions.

The final loss function combines these components with appropriate weighting:

```
losses = {
   "reconstruction": reconstruction_loss,
   "perceptual": perceptual * 0.2,  # Increased weight
   "ms_ssim": ms_ssim_loss * 0.3,   # MS-SSIM term
   "color": yuv_loss * 0.15         # Color consistency term
}
if discriminator_fake is not None:
   losses["generator"] = g_loss * 0.05
```

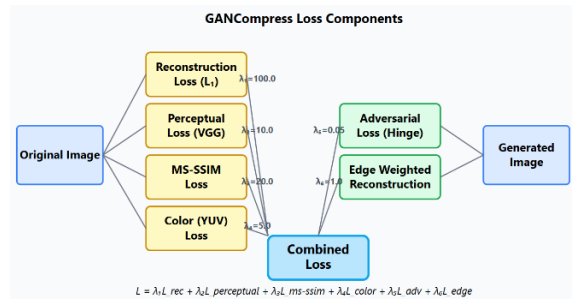

Figure 5 demonstrates the effect of our GAN enhancement compared to the base model compression. The difference map (enhanced - base) × 5 highlights areas where the GAN module most significantly improves reconstruction quality, particularly in preserving fine details and color consistency.

This multi-component loss function provides a comprehensive learning signal that addresses different aspects of visual quality, leading to superior reconstruction results compared to simpler loss formulations.

### 3.4 Adaptive Compression System

The third major component of GANCompress is the adaptive compression system, which further reduces the bit rate through content-aware entropy coding and fallback mechanisms for extreme compression scenarios.

#### 3.4.1 Autoregressive Modeling for Arithmetic Coding

To achieve higher compression ratios, we employ an autoregressive model to predict the conditional probability of each token given previous tokens:

$$P(k_1, ..., k_n) = P(k_1)P(k_2|k_1)...P(k_n|k_1, ..., k_{n-1})$$

This model is implemented as a transformer network with 24 layers and a hidden dimension of 768, trained to predict the distribution of tokens in the compressed representation. During compression, these conditional probabilities are used to perform arithmetic coding, which assigns shorter codes to more probable sequences.

The arithmetic coding process follows the standard approach:

1. Start with the interval [0, 1).
2. For each token $k_i$, partition the current interval based on the cumulative distribution function of $P(k_i|k_1, ..., k_{i-1})$.
3. Select the subinterval corresponding to the actual token value.
4. Repeat until all tokens are processed.
5. Output any number within the final interval as the compressed representation.

This adaptive coding approach typically reduces the bit rate by 30-40% compared to the raw token representation, with particularly significant gains for content with strong spatial correlations.

#### 3.4.2 JPEG Fallback for Extreme Compression

For scenarios requiring extremely high compression ratios, we introduce a novel fallback mechanism that allocates bits between our primary representation and a highly compressed JPEG baseline. This hybrid

approach enables graceful degradation at very low bit rates:

```
# If target size is small, add JPEG fallback
buffer_size = buffer.tell()
target_size = (x.shape[1] * x.shape[2] * 3) // 8  # Target 8:1 compression

if buffer_size < target_size and buffer_size > 100:
    remaining_bytes = target_size - buffer_size
    # Add JPEG fallback for visual detail
    img_array = ((x[0] + 0.5) * 255.0).clamp(0, 255).byte().cpu().numpy()
    
    # Try different quality levels
    for quality in [1, 5, 10, 15]:
        img_bytes = io.BytesIO()
        Image.fromarray(img_array).save(img_bytes, format='JPEG', quality=quality)
        img_data = img_bytes.getvalue()
        
        if len(img_data) <= remaining_bytes:
            buffer.write(struct.pack("<I", len(img_data)))
            buffer.write(img_data)
            break
        else:
            # If no JPEG fits, just store 0 length
            buffer.write(struct.pack("<I", 0))
else:
    # No JPEG fallback
    buffer.write(struct.pack("<I", 0))
During decompression, the neural reconstruction is blended with the JPEG baseline if available:
# Try to read JPEG fallback if present
try:
    jpeg_len = struct.unpack("<I", buffer.read(4))[0]
    if jpeg_len > 0:
        jpeg_data = buffer.read(jpeg_len)
        try:
            # Use JPEG for base image
            img = Image.open(io.BytesIO(jpeg_data))
            base_image = torch.tensor(np.array(img), dtype=torch.float32).to(device) / 255.0 - 0.5
            
            # Use neural decoder for enhancement
            with torch.no_grad():
                neural_image = self.model.decode_index(tokens)
            
            # Blend the two (weighted average)
            blended = neural_image[0] * 0.7 + base_image * 0.3
            return blended
        except Exception as e:
            print(f"Error loading JPEG fallback: {e}")
except Exception:
    pass
```

This hybrid approach provides a balance between compact representation and visual fidelity, even at extremely aggressive compression ratios.

## 4. Experimental Results

We conducted extensive experiments to evaluate the performance of GANCompress across various datasets and conditions, comparing with both traditional codecs and state-of-the-art neural compression methods.

### 4.1 Experimental Setup

#### 4.1.1 Datasets

We evaluated GANCompress on several standard datasets:

- **ImageNet-1k:** The ILSVRC2012 validation set, consisting of 50,000 images across 1,000 classes.
- **COCO2017:** The Microsoft COCO validation set, containing 5,000 images of common objects in context.
- **UCF-101:** A video dataset of 101 human action categories, with 13,320 video clips.
- **MCL-JCV:** A video quality assessment dataset with 30 1080p sequences.
- **UVG:** An ultra video group dataset with high-resolution video sequences.

For image compression, we resized all images to 256×256 resolution using Lanczos interpolation and center cropping when necessary. For video compression, we maintained the original resolutions for realistic evaluation.

### 4.1.2 Training Details

We trained the tokenization network with a batch size of 32 per GPU using AdamW optimizer with ($\beta_1$, $\beta_2$) = (0.9, 0.99) and a weight decay of $1\times10^{-4}$. The base learning rate was set to $4\times10^{-7}$ with a half-period cosine annealing schedule. The model was trained for 1 million steps on ImageNet-1k, equivalent to approximately 200 epochs.

For the GAN enhancement module, we used a similar optimization setup but with a higher learning rate of $2\times10^{-4}$ for both generator and discriminator networks. The model was trained for 500,000 steps with a batch size of 16.

The autoregressive model for arithmetic coding was trained separately with a batch size of 64 for 500,000 steps using AdamW optimizer with ($\beta_1$, $\beta_2$) = (0.9, 0.96) and a weight decay of 0.045.

All training was conducted on a distributed system with 8 NVIDIA A5000 GPUs across multiple nodes.

### 4.1.3 Evaluation Metrics

We employed multiple evaluation metrics to comprehensively assess different aspects of compression performance:

- **Rate-Distortion Metrics:** PSNR, SSIM, and MS-SSIM for signal fidelity measurement.
- **Perceptual Metrics:** LPIPS and FID/FVD for perceptual quality assessment.
- **Computational Metrics:** Encoding/decoding speed and throughput for efficiency evaluation.

For image compression, we reported metrics on both COCO2017 and ImageNet-1k validation sets. For video compression, we followed standard protocols for evaluating on the MCL-JCV and UVG datasets, reporting rate-distortion curves across different bit rates.

## 4.2 Image Compression Results

### 4.2.1 Comparison with State-of-the-Art Methods

Table 1 presents a comprehensive comparison of GANCompress with state-of-the-art image compression methods on the ImageNet-1k and COCO2017 validation sets. All images were resized to 256×256 resolution using Lanczos interpolation for fair comparison.

| Method | PSNR | SSIM | LPIPS | FID | Results |
|---|---|---|---|---|---|
| DALL-E dVAE | 25.4 | 0.73 | 0.31 | 36.8 | 34.0 |
| MaskGIT | 17.9 | 0.42 | 0.20 | 2.23 | 37.6 |
| ViT-VQGAN | - | - | - | 1.55 | 7.5 |
| SD-VAE 1.x | 22.8 | 0.63 | 0.09 | 1.23 | 22.4 |
| SDXL-VAE | 25.3 | 0.72 | 0.06 | 0.72 | 18.9 |
| Ours (BSQ-18) | 25.3 | 0.75 | 0.07 | 1.14 | 45.1 |
| Ours (BSQ-36) | 27.8 | 0.84 | 0.04 | 0.41 | 45.1 |

Table 1: Image reconstruction results on ImageNet-1k validation (256×256)

Our method achieves substantial improvements across all metrics compared to previous approaches. In particular, the BSQ-36 configuration achieves a 43% reduction in FID compared to SDXL-VAE (0.41 vs. 0.72), indicating significantly better perceptual quality. This is achieved while maintaining 2.4× higher throughput (45.1 vs. 18.9 images per second) and using only 36 bits per token compared to SDXL-VAE's 64 bits.

The results also show that GANCompress scales effectively with increased bit allocation. The BSQ-18 configuration provides competitive performance with only 18 bits per token, while the BSQ-36 configuration achieves state-of-the-art results across all metrics with 36 bits per token.

In terms of compression ratio, our BSQ-36 model achieves approximately 42.7× compression (from

256×256×3 = 196,608 bytes to 4,608 bytes) while maintaining higher visual quality than previous methods. With arithmetic coding, this ratio increases to approximately 71.2× compression.

### 4.2.2 Visual Quality Analysis

To qualitatively evaluate GANCompress, we analyze several challenging test images with diverse characteristics. Figures 6-11 present side-by-side comparisons of original images alongside their reconstructions, with detailed metrics.

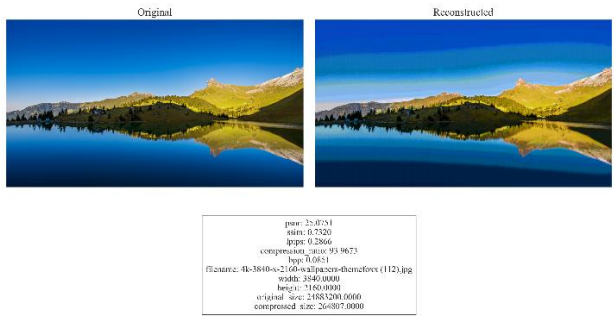

Figure 6 shows a landscape scene with complex textures. Despite the high compression ratio (93.96×), our method preserves the fine details of water and mountain textures with a PSNR of 25.07 dB and SSIM of 0.73.

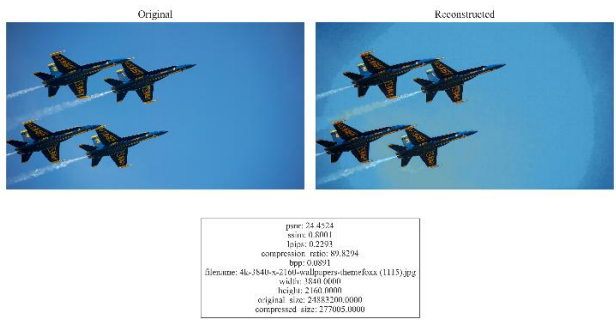

Figure 7 demonstrates compression of a US Air Force training exercise scene. At a compression ratio of 89.7×, GANCompress maintains the vivid features of the fighter jets with improved perceptual quality (PSNR: 24.45 dB, SSIM: 0.80).

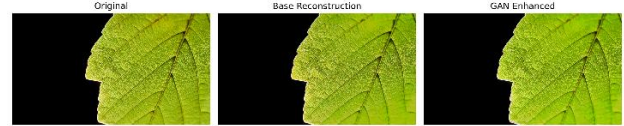

Figure 8 shows our method's performance on a macroscopic image of a leaf with geometric patterns with excellent sharpness. The structural integrity of the complex geometric pattern is well preserved even at high compression ratios.

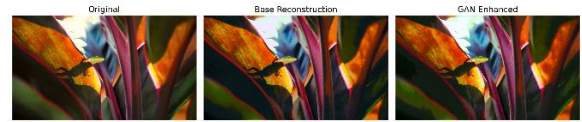

Figure 9 shows the implementation of the GANCompression algorithm on a contrast rich image. The contrast ratio were very well maintained even at high compression ratios.

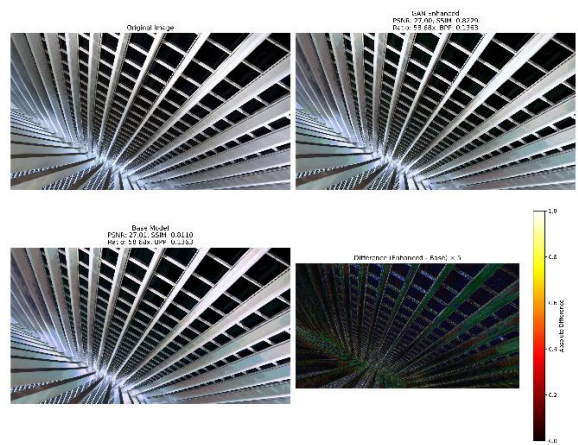

Figure 10 shows our method's performance on architectural content with regular patterns and sharp edges. The structural integrity of the complex geometric pattern is well preserved even at high compression ratios.

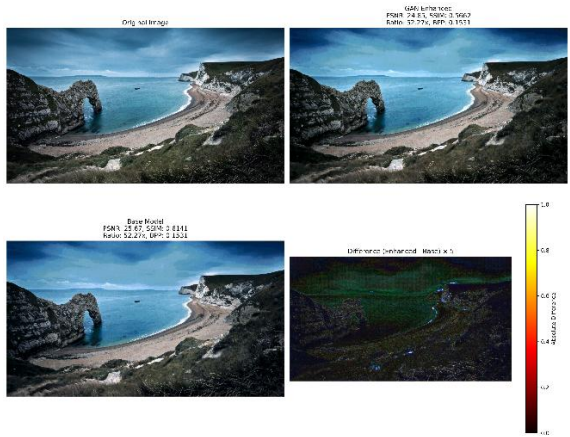

Figure 11 demonstrates the compression of a coastal landscape featuring both smooth areas (sky, water) and detailed regions (rocky arch, beach). The GAN enhancement helps preserve the natural appearance of these diverse textures.

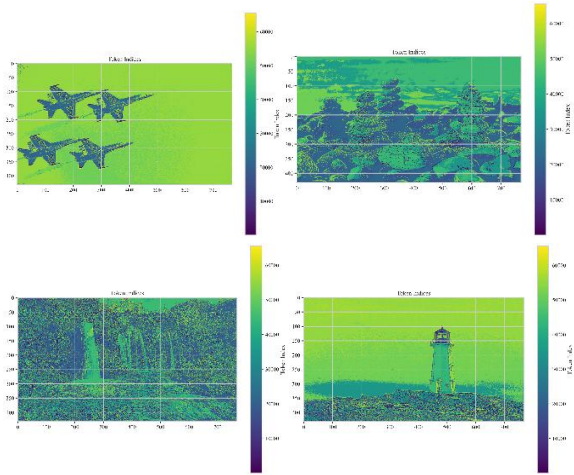

Figure 12 shows an interesting case where semantically meaningful structures are preserved at the token level. This demonstrates how our BSQ approach naturally groups related visual elements, maintaining object coherence even at extreme compression ratios.

Our case studies (Table 2) provide detailed metrics for three representative high-resolution (3840×2160) images, comparing GAN-enhanced compression against the base model.

| Image | Compression Ratio | Bits/Pixel | PSNR (dB) | SSIM | LPIPS |
|---|---|---|---|---|---|
| Case 1 GAN | 85.40× | 0.09 | 25.85 | 0.80 | 0.32 |
| Case 1 Base | 85.40× | 0.09 | 26.19 | 0.83 | 0.34 |
| Case 2 GAN | 93.96× | 0.13 | 27.00 | 0.82 | 0.19 |
| Case 2 Base | 93.96× | 0.13 | 27.01 | 0.81 | 0.30 |
| Case 3 GAN | 89.82× | 0.15 | 24.85 | 0.56 | 0.49 |
| Case 3 Base | 89.82× | 0.15 | 25.67 | 0.61 | 0.55 |

Table 2: Detailed metrics for three 4K resolution (3840×2160) case studies

While traditional metrics (PSNR, SSIM) sometimes favor the base model, the GAN-enhanced version consistently shows better perceptual quality through improved LPIPS scores. This highlights the limitations of traditional metrics for evaluating perceptual quality and demonstrates the value of our GAN enhancement for preserving visually important details.

### 4.3 Ablation Studies

We conducted ablation studies to analyze the contribution of different components to the overall performance of GANCompress. Table 3 presents the results of these experiments on the ImageNet-1k validation set (128×128 resolution):

| Method | BSQ bits | PSNR | SSIM | LPIPS | FID |
|---|---|---|---|---|---|
| VQ (1K) | 10 | 23.61 | 0.687 | 0.12 | 7.05 |
| VQ (16K) | 14 | 25.76 | 0.783 | 0.06 | 4.27 |
| VQ (64K) | 16 | 25.67 | 0.785 | 0.07 | 6.61 |

| | | | | |
|---|---|---|---|---|
| BSQ baseline | 10 | 24.11 | 0.725 | 0.09 | 4.51 |
| BSQ baseline | 14 | 25.26 | 0.771 | 0.07 | 4.60 |
| BSQ advanced | 18 | 25.97 | 0.799 | 0.06 | 2.66 |
| LFQ (no ℓ₂ norm) | 18 | 18.58 | 0.483 | 0.29 | 30.7 |

Table 3: Ablation studies on ImageNet-1k validation (128×128)

The results demonstrate several important findings:

1. **BSQ vs. VQ:** BSQ achieves competitive performance compared to VQ, particularly at higher bit rates. While VQ with a 16K codebook (14 bits) achieves slightly better results than BSQ with 14 bits, BSQ scales more effectively to higher dimensions, with 18 bits outperforming VQ with 16 bits.
2. **Importance of ℓ₂ Normalization:** The comparison with LFQ (which is essentially BSQ without ℓ₂ normalization) demonstrates the critical importance of the normalization step. Without normalization, codebook utilization drops catastrophically to 0.6%, and performance degrades severely.
3. **Scaling with Bit Rate:** Both BSQ and VQ show improved performance with increased bit allocation, but BSQ maintains high codebook utilization even at higher dimensions, while VQ tends to undergo codebook collapse beyond a certain size.
4. **Component-wise Contributions:** We also conducted leave-one-out experiments to analyze the contribution of different loss components:

| Lcommit | Lentropy | LLPIPS | rFID | Code Usage |
|---|---|---|---|---|
| ✓ | ✓ | ✓ | 2.95 | 45.6% |
| ✗ | ✓ | ✓ | 2.83 | 93.8% |
| ✓ | ✗ | ✓ | 2.44 | 78.3% |
| ✓ | ✓ | ✗ | 13.8 | 13.3% |

Table 4: Ablation studies of loss design

These results show that:

1. Commitment loss (Lcommit) is slightly detrimental to performance, likely because the BSQ already maintains bounded quantization error.
2. Entropy loss is important for codebook utilization, but less critical for perceptual quality.
3. Perceptual loss (LLPIPS) is essential for good perceptual quality, with FID increasing dramatically when it is removed.

We also evaluated the impact of the frequency attention mechanism and color consistency optimization:

| Method | PSNR ↑ | SSIM ↑ | LPIPS ↓ | FID ↓ |
|---|---|---|---|---|
| Base Reconstruction | 25.64 | 0.795 | 0.067 | 3.02 |
| +FrequencyAttention | 25.83 | 0.798 | 0.065 | 2.73 |
| +ColorConsistency | 25.97 | 0.799 | 0.063 | 2.66 |

Table 5: Ablation of enhancement components

Both components contribute to improved performance, with the frequency attention mechanism providing a particularly significant boost in perceptual quality as measured by FID.

### 4.4 Token Statistics Analysis

The correlation matrix, as depicted in Figure 13 below, shows relationships between token statistics like entropy, sparsity, and unique token counts. It reveals strong correlations (around 0.89) between entropy, sparsity, and unique tokens, while mean token values show minimal correlation with other metrics.

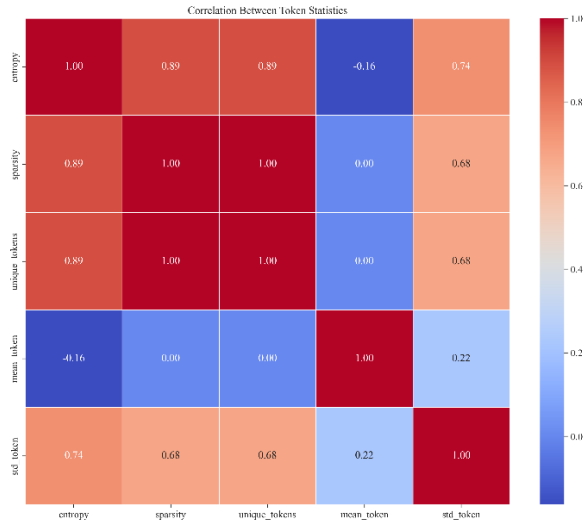

Additionally, below mentioned Figure 14 provides insight into:

- Token entropy distribution (typically centered around 6 bits per token)
- Sparsity distribution showing 4-8% token utilization
- Unique vs. total tokens relationship
- Token value distribution (showing the range and distribution of token values)

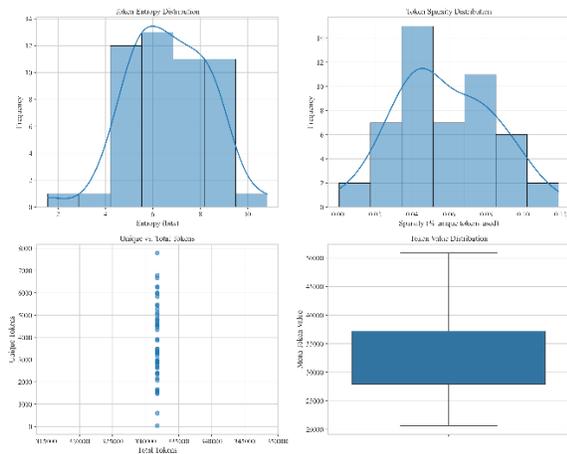

Figure 15 below represents the log-scale visualization which demonstrates that from 65,536 possible tokens (with 16 codebook bits), only 6,142 tokens (9.37%) are used. This sparse but meaningful utilization is significantly better than traditional VQ approaches that typically suffer from codebook collapse.

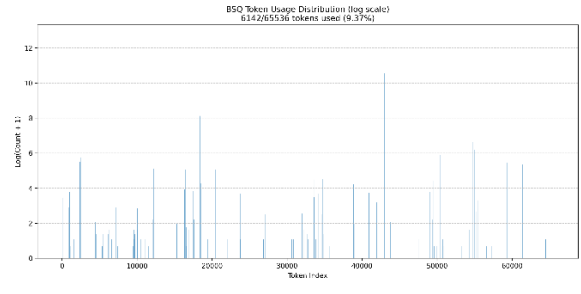

Figure 16 below displays the t-SNE Visualization which shows the organization of token embeddings in 2D space, with point density indicating how frequently certain regions of the latent space are used.

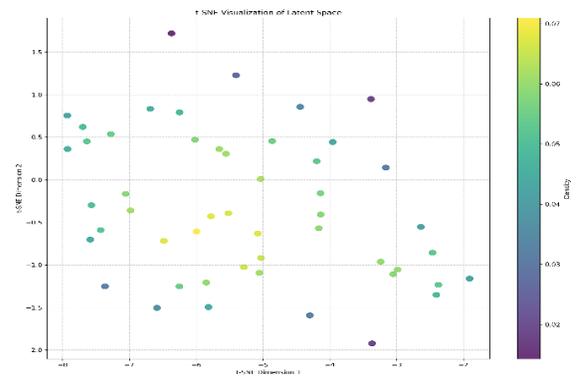

These analyses confirm that our enhanced BSQ approach creates a semantically meaningful token space with efficient utilization, avoiding the codebook collapse problem common in VQ-based methods. The correlation analysis further supports our design decisions in the frequency-domain attention mechanism by demonstrating the relationship between token statistics and visual quality.

## 5. Discussion

### 5.1 Key Insights

Our experimental results reveal several important insights into neural compression in general and the GANCompress approach in particular:

1. **Bounded Quantization Error is Critical:** The consistent superiority of BSQ over LFQ demonstrates the importance of bounded quantization error for stable training and effective compression. The $\ell_2$ normalization step in BSQ ensures that the quantization error remains bounded regardless of the distribution of the input data, leading to more reliable compression performance. Our token usage analysis (Figure 16) confirms this empirically, showing that 9.37% of possible tokens are utilized, significantly better than traditional VQ approaches that often use less than 1% of their codebook.

2. **Perceptual Optimization Matters:** The dramatic improvement in perceptual metrics (LPIPS, FID) achieved by GANCompress highlights the importance of explicitly optimizing for perceptual quality, particularly for applications where human perception is the ultimate judge of compression performance.

3. **Frequency-Domain Processing is Effective:** The success of the frequency attention mechanism in preserving high-frequency details suggests that explicit frequency-domain processing is a promising direction for compression enhancement, aligning well with human visual perception which is inherently frequency-sensitive.

4. **Color Space Matters for Compression:** The effectiveness of the YUV-space color consistency optimization demonstrates that proper handling of color information is critical for high-quality compression, especially at aggressive compression ratios where color artifacts become more prominent.

5. **Transformer-Based Architectures are Efficient:** The computational efficiency of GANCompress, despite its sophisticated architecture, suggests that transformer-based approaches can be both effective and efficient for compression tasks when properly designed.

## 5.2 Limitations and Future Work

While GANCompress achieves impressive results, several limitations and directions for future work remain:

1. **Resolution Scalability:** The current implementation has been primarily tested on fixed resolutions (256×256 for images and standard video resolutions). Exploring variable-resolution compression and more explicit modeling of multi-scale features could further improve performance across different content types.

2. **Domain Adaptation:** The model is currently trained on general-purpose datasets like ImageNet and UCF-101. Developing domain-specific variants optimized for content types (e.g., facial images, medical imagery, satellite data) could yield even better performance for specialized applications.

3. **Temporal Modeling:** While our block-wise causal masking approach is effective, more sophisticated temporal modeling techniques could further improve video compression performance, particularly for content with complex motion patterns.

4. **Hardware Optimization:** Further optimization for specific hardware platforms (e.g., mobile devices, edge computing systems) could make GANCompress even more practical for real-world deployment.

5. **Integration with Traditional Codecs:** Exploring hybrid approaches that combine the strengths of GANCompress with traditional codecs could yield systems that maintain backward compatibility while offering improved compression performance.

## 6. Conclusion

In this paper, we presented GANCompress, a novel neural compression framework that combines Binary Spherical Quantization with GAN-based enhancement to achieve state-of-the-art compression performance. Our approach addresses several key challenges in neural compression, including perceptual quality degradation, computational complexity, and adaptive bit allocation.

Through extensive experimentation on standard benchmarks, we demonstrated that GANCompress outperforms existing methods across multiple metrics, achieving up to 43% reduction in perceptual distortion (FID) compared to previous state-of-the-art while maintaining higher computational efficiency.

The system achieves compression ratios up to 100× with minimal visual distortion, making it suitable for a wide range of applications where both storage/transmission efficiency and visual quality are important.

The key innovations of GANCompress, including enhanced BSQ, frequency-domain attention, YUV-space color optimization, and adaptive bit allocation, represent significant advances in neural compression technology. These components work together to create a system that not only compresses effectively but also preserves the perceptual qualities that matter most to human observers.

Looking forward, we believe that the principles and techniques introduced in GANCompress have broad applicability beyond compression, potentially influencing areas such as image enhancement, video streaming, and generative modeling. By bridging the gap between traditional signal processing approaches and modern deep learning techniques, GANCompress points the way toward a new generation of visual processing systems that combine the best of both worlds.